**Understanding Bulk Defects in Topological Insulators from Nuclear-Spin Interactions**

*Dimitrios Koumoulis, Belinda Leung, Thomas C. Chasapis, Robert Taylor, Daniel King Jr., Mercouri G. Kanatzidis, Louis S. Bouchard\**

Dr. D. Koumoulis, B. Leung, Dr. R. Taylor, D. King Jr., Prof. L. Bouchard
Department of Chemistry and Biochemistry, California NanoSystems Institute, 607 Charles Young Drive East, University of California, Los Angeles, 90095, USA
Email: bouchard@chem.ucla.edu
Dr. T.C. Chasapis, Prof. M.G. Kanatzidis
Department of Chemistry and Biochemistry, Norwestern University, Evanston, IL, USA
Email: m-kanatzidis@northwestern.edu

Non-invasive local probes are needed to characterize bulk defects in binary and ternary chalcogenides. These defects contribute to the non-ideal behavior of topological insulators. We have studied bulk electronic properties via $^{125}$Te NMR in $Bi_2Te_3$, $Sb_2Te_3$, $Bi_{0.5}Sb_{1.5}Te_3$, $Bi_2Te_2Se$ and $Bi_2Te_2S$. A distribution of defects gives rise to asymmetry in the powder lineshapes. We show how the Knight shift, line shape and spin-lattice relaxation report on carrier density, spin-orbit coupling and phase separation in the bulk. The present study confirms that the ordered ternary compound $Bi_2Te_2Se$ is the best TI candidate material at the present time. Our results, which are in good agreement with transport and ARPES studies, help establish the NMR probe as a valuable method to characterize the bulk properties of these materials.

**1. Introduction**

The band structure of $V_2VI_3$ layer-type semiconductors offers a desirable platform for thermoelectric (TE) and topological insulating (TI) properties.[1-39] Prototype TE and TI compounds of interest in applications include antimony telluride ($Sb_2Te_3$) and bismuth telluride ($Bi_2Te_3$).[7-11,20-22,26-32] These binary chalcogenides have been extensively studied due to their large thermo-electric power and applications as solid state power generators and refrigerators (Peltier, Seebeck and Thomson devices).[20-22,26-29] Recently, an effort aimed at improving the TE efficiency (figure of merit parameter, *ZT*) concluded that $Bi_2Te_3$-$Sb_2Te_3$ solid solutions yield improved TE materials, according to the *ZT* parameter, which is maximized in the case of $Bi_{0.5}Sb_{1.5}Te_3$.[7,17-20,24-29] Other ternary telluride compounds in the same class as these layered materials, $Bi_2Te_2Se$ and $Bi_2Te_2S$, have shown the potential of tuning and enhancing transport properties.[1-6,30,32]

The discovery of topological insulators has changed the way we look at chalcogenides.[11-16] An ideal TI has a bulk interior characterized by an insulating band gap, while the boundary exhibits gapless Dirac-like edge (2D-TIs) or surface states (3D-TIs).[11-16,36-43] Experimental realizations of TIs to date have used narrow band gap materials with strong spin-orbit coupling (SOC) featuring a helical surface band structure with spin locked to momentum. The observation of metallic surface states often requires minimal bulk defects or compensation for the free carriers.[1-6,11,14,37,38-53] Novel properties have been predicted, such as quantum-Hall-like behavior with quantized conductance of charge and spin in the absence of a magnetic field, and dissipation-less quantum Hall states with time-reversal symmetry (TRS).[11-16] First-principles calculations[3,15,16,33,36-40,48,49,51] and Angle Resolved Photoemission Spectroscopy (ARPES)[5,11,12,39,40,41,51] results have yielded energy band structures of most binary and ternary chalcogenides.

The characterization of TI states requires sensitive techniques to probe metallicity of the surface state, because the surface is where the interesting properties are found.[11] To date, electrical transport measurements, scanning tunneling microscopy (STM) and ARPES have been the main workhorses for the study of surface states.[1-6,11,13,40,41,42,43] These techniques work best with high-quality thin films (<20 nm)[40-43] or large single crystals and at low temperatures (<30 K).[1-6,11,39,42,43] There is, however, a need for characterizing materials at room temperature or materials of suboptimal quality. To this end, nuclear magnetic resonance (NMR) has been proposed as a tool to characterize TI properties.[54-56] NMR is a non-invasive local probe of magnetism and electronic wavefunction in narrow-gap semiconductors.[57-61] It has been used to study the bulk semiconductor properties of chalcogenides,[47,54-56] as well as the properties of TI nanoparticles.[54] Worthwhile aspects of the NMR readout include: (i) potential operation at higher temperatures and ambient conditions, (ii) ability to probe lower quality or amorphous materials, (iii) study of "granular" varieties and (iv) materials with a large number of bulk defects with topologically protected gapless modes.[54-56,59,61] Magnetic resonance techniques could serve as a complementary tool to characterize materials which are not suitable for study by conventional methods (transport, ARPES, STM). Of great interest is the development of local probes of the material's bulk region, which is difficult to interrogate independently from the surface states.[52,53] Here we present a comparative NMR study of the bulk states of multiple TI materials and have relate the results to existing transport and ARPES studies from the literature. The present study confirms that the undesired presence of defects in the bulk state are directly reflected in the NMR parameters.

## 2. Material Properties

Bi$_2$Te$_3$, Sb$_2$Te$_3$ and Bi$_2$Se$_3$ are the simplest members of the Tetradymite family, which are crystallized in the primitive rhombohedral structure belonging to the $D_{3d}^5$ ($R\overline{3}m$) space group. Their structure is arranged in quintuple layers stacked upon each other along the c-axis, separated by van der Waals gaps. For Bi$_2$Te$_3$ and Sb$_2$Te$_3$, for example, the layers are arranged in the following way: Te(2)-Bi-Te(1)-Bi-Te(2) and Te(2)-Sb-Te(1)-Sb-Te(2), respectively, where the labels (1) and (2) denote crystallographically different sites. The Te(2) sites refer to the outer planes of the quintuples which are exposed to the van der Waals gaps and host the topological surface states.[6-9,16-30]

The type of conduction in the Tetradymite family of TIs is strongly related to the type of structural defects present (anion vacancies or antisite).[1-14,16,22,25,28,32,35] Bi$_2$Te$_3$ crystals grown from stoichiometric melts are *p*-type due to the presence of large number of Bi$_{Te}$ antisite defects. The type of conduction may be adjusted through fine tuning of the non-stoichiometry of the Bi/Te ratio, i.e., to dope the system by generating antisite defects of $Bi_{Te}$ (*p*-type) or $Te_{Bi}$ (*n*-type[18]). The most common defects in $Sb_2Te_3$ are the $Sb_{Te}$ antisite defects, giving rise to large *p*-type conduction.[6,10,21,23] The presence of weak bond polarity[3,6,9,23,25,26,29,32,35,39,62-64] (difference in electronegativity) is a key parameter that explains the formation of antisite defects in these materials. Even a small discrepancy in stoichiometry or dopants mixed into the Tetradymite matrix can affect the initial polarity of the bonds, leading to antisite defects.[8,10,13,20,45,62-65] $Bi_2Se_3$, as prepared, is observed to be *n*-type owing to a large number of Se vacancies formed at the chalcogen outer layers due to evaporation.[6,19] Recently, $Bi_{Se}$ antisite and partial $Bi_2$-layer intercalation defects were also identified in stoichiometric and non-stoichiometric $Bi_2Se_3$.[14]

A common route to optimize TI material properties involves alloying of the binary starting materials. Alloying offers two degrees of freedom: tuning the carrier density, i.e. the position of the Fermi level, and/or moving the Dirac point relative its position in the parent binary compound. Given the similarities in crystal structure and lattice constants, ternary compounds of the type $(Bi_{1-x}Sb_x)Te_3$ were proposed to tailor the Dirac point and provide charge compensation by adjusting the Bi/Sb stoichiometry. $Bi_2Te_3$ has an indirect band gap equal to 0.16 eV and its Dirac point lies in the valence band.[5,18,22,26,27,31,45,47,56] **Figure 1**a compares ARPES band structures of (1) $Bi_2Te_3$, (3) $Bi_{0.5}Sb_{1.5}Te_3$ and (4) $Sb_2Te_3$, along the symmetry paths K-Γ-M and M-Γ-M in the first Brillouin zone, where the Dirac point is clearly more exposed in the case of the solid solution (3) relative to the parent binary $Bi_2Te_3$.[5,40,36,37] In $Bi_{0.5}Sb_{1.5}Te_3$ (direct band gap, 0.2 eV), the Sb dopant alters the initial density of defects but keeps the bulk carrier density constant without affecting the TI surface states.[16,20,21,24,25,40,41] Recently, *in situ* ARPES results have shown that as the Bi/Sb ratio approaches 0.25/0.75, the Fermi energy shifts to the bulk valence band (BVB) leading to a hybrid state consisting of the surface state band (SSB) and the BVB pocket.[40,41,48]

Alloying two parts of $Bi_2Te_3$ for each part of $Bi_2Se_3$ forms the structurally ordered $Bi_2Te_2Se$ phase [i.e., 66.7% $Bi_2Te_3$ with 33.3% $Bi_2Se_3$], where Se atoms being trapped between two Bi atoms and the basic quintuple-layer unit being Te-Bi-Se-Bi-Te. The phase was found with *n*-type conduction and high bulk resistivity at low temperatures, 1 Ω·cm.[4-5] The low carrier concentration originates from a compensation mechanism between $Bi_{Te}$ antisites and Te vacancies which dope holes and electrons respectively, while the Se vacancies are suppressed relative to the $Bi_2Se_3$ because the Se atoms trapped between two Bi atoms are less exposed to evaporation.[1] Considering that $Bi_2Se_3$ has the Dirac point of the surface states at an energy well separated from the energies of the top of the bulk valence band and the bottom of the bulk conduction band, the Dirac point energy of the $Bi_2Te_2Se$ shown in (2) of **Figure 1**a, appears at a higher energy than in $Bi_2Te_3$ [2,4-6,40] Although $Bi_2Te_2Se$ is one of the most important compositions for studying the surface quantum transport in a topological insulator,[1-6] its Dirac point appears to be near the energy of the top of the valence band, suggesting that the bulk states will interfere with surface states conductivity. By replacing the element Se with the more electronegative S in $Bi_2Te_2Se$, the absolute energy of the valence band is expected to decrease below the Dirac point.[6] Recently, a S-rich $Bi_2Te_{2-x}S_{1+x}$ composition has been the subject of investigation as the host for topological surface states. In particular, the *n*-type $Bi_2Te_{1.6}S_{1.4}$ which incorporates both S and Te in its outer chalcogen layers has shown that the layers supporting the surface states are randomly corrugated on the atomic scale, while partial substitution of Sb for Bi yielded a high resistivity material with well isolated Dirac point[5] According to ARPES results and *ab initio* calculations,[1-6,37,39] $Bi_2Te_2S$ as well as $Bi_2Te_2Se$ exhibit striking similarities with $Bi_2Te_3$. Defects are found both on the surface and the bulk of these TIs.[1-6,39]

The defect density in the all Tetradymite TIs is strongly dependent on growth method used for synthesis and the form of the final material (single crystals, thin films, nanocrystals, etc.). These factors, in turn, determine the actual carrier concentrations.[1-6,22-26,40,41] ARPES and

STM measurements are excellent probes of TI surface defects. However, they cannot easily probe antisite defects and vacancies in the bulk, whose properties are different than on the surface. In any case, this study explores NMR techniques as a complementary characterization tool to overcome some of the limitations associated with probing the bulk.

3. Results and Discussion

**Figure 1**b shows the powder XRD patterns of the studied compositions. The patterns for all the samples can be well indexed with the rhombohedral structure where the 015 reflection of the hexagonal unit cell is the strongest, confirming that the ternaries compositions maintain the same crystal structure as the binaries counterparts. Based on the PXRD spectra, the materials are single phase materials except for the $Bi_2Te_2S$ which was found phase separated. This is actually supported by the comparison of the respective PXRD pattern with the pure $Bi_2Te_3$ and $Bi_2Te_2S$ reflections shown in **Figure 1**c. The experimental pattern is well described by taking into account the contribution of two phases. As can be seen from Figure 1c, the PXRD peaks at ~28° and ~29°, attributed to the 015 reflections of the two hexagonal phases, are blue shifted relative to the pure compounds indicating that the material may be regarded as being composed by a S-poor and a S-rich $Bi_2Te_3$ compositions, with the latter having the strongest contribution.

The lattice constants and the volume of the hexagonal unit cell are obtained after refinement of the respective PXRD spectra and are tabulated in **Table 1**. The values lie close to the literature values. For the phase separated composition the refinement based on the two pure phases yielded two pairs of in-plane, out-of plane lattice constants; 4.349Å, 30.487Å, 4.226Å, and 29.500Å. The first pair is attributed to the S-poor constituent. Based on the lattice constants of the pure $Bi_2Te_2S$, 4.239Å, 29.576Å, we may conclude that the composition of the second phase is $Bi_2Te_{2+\delta}S_{1-\delta}$. Regarding the carrier concentrations it is evident from **Table 1** that the $Bi_2Te_2Se$ composition shows the lowest value in the range of $10^{18}$ cm$^{-3}$ supporting the idea that the ordered phase is the most promising TI.

$^{125}$Te NMR spectra of $Bi_2Te_3$, $Bi_{0.5}Sb_{1.5}Te_3$, $Sb_2Te_3$, $Bi_2Te_2Se$, and $Bi_2Te_2S$ at temperatures in the range 170-423 K are shown in **Figure 2**. The resonance line shapes of Sb-TIs are clearly asymmetric, especially in the case of $Bi_{0.5}Sb_{1.5}Te_3$. Such asymmetrical features arise predominantly from interactions of nuclear moments with conduction band carriers.[54,55,59,60,61] In prior studies of $Bi_2Se_3$ and $Bi_2Te_3$, the asymmetric resonance lines were attributed to an inhomogeneous distribution of defects in a low symmetry rhombohedral space group environment yielding different charge carrier concentrations in different crystalline regions, resulting in a range of Knight shifts.[54,55] A spectral model for the Sb-TI samples and the phase separated $Bi_2Te_2S$ is a sum of two Lorentzians. For $Bi_2Te_3$ and $Bi_2Te_2Se$, the line shapes are more symmetric and are well described by a single Lorentzian line.

From the data of **Figure 2** we extracted center frequency and linewidth of each Lorentzian line (for both central and shoulder peaks) and summarized the results as function of temperature in **Figure 3**. We observe the following features: 1) the NMR shift of central peaks of $Bi_2Te_3$ and $Bi_{0.5}Sb_{1.5}Te_3$ become more negative (lower frequency) with decreasing temperature, except for $Sb_2Te_3$, $Bi_2Te_2Se$ and $Bi_2Te_2S$ which remain essentially constant within experimental uncertainty. 2) In all cases, the NMR shift of the shoulder peaks exhibits a strong temperature dependence. Furthermore, this temperature dependence of the NMR shift is more pronounced at low temperatures in the case of $Bi_{0.5}Sb_{1.5}Te_3$, suggesting an increase in carrier density. Knight shifts in narrow-band multi-valley semiconductors (Equation 1) have been explained previously in terms of SOC.[44-47,55-61,68] One of the defining features of such Knight shifts is their temperature dependence, which gravitates toward more negative frequencies as temperature decreases, owing to the temperature dependence of the energy gap.[44-47,55-59] According to the theory,[44-46] the Knight shift can be expressed as,

$$K = \zeta \frac{8\pi}{3} \gamma_e^2 h^2 \langle |u_k(0)|^2 \rangle_{E_0} \frac{n_e}{kT}, \quad (1)$$

where $\gamma_e$ is the electron magnetogyric ratio, $h$ is Planck's constant, $n_e$ is the carrier density, $\langle |u_k(0)|^2 \rangle_{E_0}$ is the free electron density near the bottom of the conduction band for electrons or near the top of the valence band for holes and $\zeta$ is a numerical factor close to unity which depends on the origin and strength of the hyperfine interaction (Fermi contact, dipolar or orbital). The observed $^{125}$Te Knight shift unveils different bulk electronic states and density of states at the Fermi level in the present TIs, which we attribute to large differences in defect and vacancy concentration. Transport, ARPES and structural studies in the same materials[1-6,11,16,37,41-43,69,70] are in agreement and support the above NMR Knight shift behavior.

A plot of linewidth as function of temperature is shown in **Figure 3b**. As mentioned previously, the linewidth in these samples reflects an inhomogeneous distribution of native defects with an increased carrier density having different Knight shifts.[47,54-56,58-61] A previous NMR study on Bi$_2$Se$_3$ [56] reported a narrower spectral linewidth in samples with lower defect concentration, which is in agreement with our study of all present TIs. We identify that the temperature dependence of the linewidth in all the samples is different (**Figure 3**c). In addition, we note that as the temperature decreases, the linewidth for the ternary compound Bi$_{0.5}$Sb$_{1.5}$Te$_3$ is larger compared to the binary TIs. The Bi$_{0.5}$Sb$_{1.5}$Te$_3$ linewidth is more than 300 ppm larger than that of Bi$_2$Te$_3$ near 215 K. Although the five materials share the same rhombohedral structure, a different defect level gives rise to the observed difference of the temperature dependent linewidths. This is due to the increased broadening via spin-flip scattering events by the charge carriers in the bulk,[54-61] which are found to be at a higher concentration in the case of Bi$_{0.5}$Sb$_{1.5}$Te$_3$ due to the higher density of defects.[8-11,64,69,70] Indeed, recent $^{209}$Bi NMR results on Bi$_2$Se$_3$ with different defect levels show a significant difference in linewidths, which was attributed to different amounts of charge carriers.[56] Another interesting feature we observe with all these TIs is that the linewidth of the shoulder peak decreases with increasing temperature, as would be expected due the motional narrowing of the lineshapes accompanied with a lack of annealing of the native defects at higher temperatures.[54,55,58-61,68] Dislocations, strains and vacancies in the lattice can also enhance the observed line broadening, e.g., due to the substitution of Sb or S in the $Bi_2Te_3$ matrix, especially in the case of the solid-solution $Bi_{0.5}Sb_{1.5}Te_3$[10,17,24,26,28,33,35] and $Bi_2Te_2S$ (a phase separated material).[5,6,71,72]

$^{125}$Te spin-lattice relaxation in materials which are dominated by structural defects is best described with a stretched exponential model (Kohlrausch function) [54,55,57,59,61,60,73]

$$M(t) = M_0(1 - \exp\left(-\frac{t}{T_1}\right)^\beta), \quad (2)$$

where $T_1$ is the spin-lattice relaxation time and $\beta$ is the Kohlrausch exponent. For $Bi_{0.5}Sb_{1.5}Te_3$ and $Sb_2Te_3$, $\beta=0.5$ and $\beta=0.9$, respectively, provided a good fit, whereas $\beta=1$ (single exponential) gave a better fit for $Bi_2Te_3$. With $Bi_2Te_2S$ and $Bi_2Te_2Se$, $\beta=0.7$ provided the better fit. Typical $T_1$ saturation recovery curves for Sb-TIs, $Bi_2Te_2S$ and $Bi_2Te_2Se$ are shown in **Figure 4** (see also Reference 55 for $Bi_2Te_3$). As shown in **Figure 4** insets for all cases the $\beta$ has a temperature independent behavior and thus the $T_1$ does not change its overall behavior across the entire temperature range since there is a constant underlying distribution in $T_1$. In order to avoid unnecessary experimental scattering in $T_1$ process, we return back and fit again all the saturation recovery data to Equation (2) using the best fixed value of $\beta$. The stretched exponential has been used for many decades to model NMR relaxation characterized by a distribution of relaxation rates in materials such as semiconductors. The $\beta$ exponent is a measure of the width of the $1/T_1$ distribution.[57] Recently, another detailed $^{209}$Bi NMR study by Nisson *et al.* on single crystal TI materials ($Bi_2Se_3$, $Bi_2Te_2Se$) conclude that the $T_1$ recovery data fits better to the stretched exponential model.[57]

The natural logarithm of the spin-lattice relaxation rate as a function of the inverse temperature, obtained from the fitting results, is shown in **Figure 5**a. The binary chalcogenides follow a two-channel relaxation process, whereas the results of $Bi_{0.5}Sb_{1.5}Te_3$ follow a single relaxation mechanism across the entire temperature region (170-423 K). In the case of $Bi_2Te_3$ and $Sb_2Te_3$, the first relaxation mechanism occurs in the low temperature regime (<270 K) and follows a thermal activation process with an activation energy of about 2.26 kJ/mol (23 meV) for $Bi_2Te_3$ and 0.68 kJ/mol (7 meV) in case of $Sb_2Te_3$. Above 270 K, another relaxation mechanism becomes apparent, which is characterized by an activation energy of 8.44 kJ/mol (87 meV) for $Bi_2Te_3$ and 5.13 kJ/mol (53 meV) for $Sb_2Te_3$. This high temperature process is typical of semiconductors with nuclear spins that interact with thermally activated charge carriers. The $Bi_{0.5}Sb_{1.5}Te_3$, on the other hand, exhibits an Arrhenius behavior across the entire temperature region with activation energy equal to 4.89 kJ/mol (51 meV). While these values for the activation energy (reflect inter-band excitations and not to be confused with the band gap between the valence and conduction band) should not be taken too literally given the narrow range of temperatures investigated, the analysis does allow for the identification of distinct relaxation mechanisms. We also note that there is no significant energy difference at the high temperature regime between $Sb_2Te_3$ and $Bi_{0.5}Sb_{1.5}Te_3$. However, the low temperature relaxation mechanism does not present itself in the relaxation process for $Bi_{0.5}Sb_{1.5}Te_3$. A further analysis by a power-law fit ($1/T_1 \propto T^n$) of relaxation rates (**Figure 5**d) confirms the existence of a slow temperature dependence of the form $1/T_1 \propto T^{0.3}$ for $Sb_2Te_3$ and a $1/T_1 \propto T^{1.2}$ for $Bi_2Te_3$ below 270 K, whereas a Raman-type process is likely to govern all the samples above 270 K, as evidenced by the $T^{n \geq 2}$ dependence of two-phonon relaxation processes. The Raman relaxation mechanism involving the interactions between nuclear spins and lattice vibrations is also active in the case of spin-*1/2* nuclei (such as $^{125}Te$, $^{207}Pb$, $^{77}Se$, $^{119}Sn$, etc.). Especially in case of $^{125}Te$ and $^{77}Se$ NMR, the existence of Raman process has been known since 1970s [78-81]. The above view is supported by the observation of a temperature-dependent $T_1$ relaxation process proceeding by the inelastic scattering of phonons by the spin (Raman mechanism). Recently, an unusual power-law dependence of the electrical resistivity and Raman measurements in the temperature range 20-270 K has been observed in the case of binary TIs ($Sb_2Te_3$ and $Bi_2Te_3$) accompanied by an anomalous thermal expansion in the temperature range 221-228 K.[65,75-77]

The $Bi_2Te_2Se$ and $Bi_2Te_2S$ behave much like $Bi_2Te_3$ throughout the entire temperature range, but due to lower bulk conductivities, exhibit higher activation energies. These NMR results are in agreement with *ab-initio*, magneto-transport and ARPES results.[1-6] $Bi_2Te_2Se$ and $Bi_2Te_2S$ feature a higher bulk insulating behavior compared to $Bi_2Te_3$ and $Bi_2Se_3$ as a result of a decrease in the SOC strength and limited antisite Bi-Te and Se vacancies, which in turn, is expected to result in longer relaxation times. This is reflected in our NMR results of **Figure 5**. Additionally, based on recent *ab initio* calculations,[37] the SOC constant in $Bi_2Te_2Se$ is two times weaker than in $Bi_2Te_2S$, $Bi_2Te_3$ and $Bi_2Se_3$. The SOC contribution to the spin-lattice relaxation in TI materials has also been observed in a previous NMR study on $Bi_2Se_3$ and $Bi_2Te_3$.[55]

It is interesting to compare the above results on TIs with a recent $^{125}$Te NMR study of PbTe, a narrow-gap semiconductor (0.32 eV) which is not a TI.[61] PbTe exhibits an activation energy of approximately 15.36 kJ.mol$^{-1}$ (159 meV) (nearly three times larger than that found for all the above TIs) in the higher temperature regime, and a second relaxation mechanism at lower temperatures characterized by an activation energy of 6.05 kJ.mol$^{-1}$ (63 meV), nearly seven times larger than that found for all the above TIs. The smaller activation energy of TIs is attributed to excitations of electrons from impurity band states (defect or vacancy localized states) which lie above the valence band into the conduction band.[55] The strong temperature-dependent relaxation mechanism, especially in the case of binary TIs, is attributed to the presence of an indirect bandgap. A strong electron (or hole)-phonon coupling associated with changes in the bonding lengths[1-6,10,35,37,65, 71,72,74-77,82] (Bi-Te, Sb-Te, Te-Se,Te-S) can account for this effect, as has already been observed in the case of another chalcogenide ($MoSe_2$ [74]). In this system, thermal excitation across an indirect bandgap requires the addition of a phonon-assisted mechanism (associated with a change in crystal momentum) in the first Brillouin zone. [10,22,74]

The temperature dependence of the Korringa product[59] $T_1.T$ for all samples is shown in **Figure 5**b. A lower value of the $T_1.T$ means a higher density of states at the Fermi level ($D^2(E_F)$), since $1/T_1.T \sim D^2(E_F)$, leading to a more conductive bulk.[55-61] The lowest Korringa products across all temperatures are associated with $Bi_{0.5}Sb_{1.5}Te_3$ and $Sb_2Te_3$, which implies that $D^2(E_F)$ is higher compare to the other TIs. On the other hand, $Bi_2Te_3$ reveals a Korringa law below 200 K and a semiconducting behavior at high temperatures. A critical parameter that affects the $T_1$ properties is the SOC of the p-band electrons.[37,40,41,55] *Ab initio* calculations and ARPES results[33,36,40,41] suggest that the $Bi_{0.5}Sb_{1.5}Te_3$ compound has an increased *p*-band conductivity in bulk states compared to the binary parent compounds ($Bi_2Te_3$ and $Sb_2Te_3$).

In **Figure 5** inset, we plot the quantity $1/(T_1.T^{0.5})$ which is proportional to the charge carrier concentration in a semiconductor (*N*) [54,56,59,61,68] The $Bi_{0.5}Sb_{1.5}Te_3$ follows a linear trend which is indicative of a metallic sample across the temperature range investigated. **Figure 5**c shows that the $Bi_{0.5}Sb_{1.5}Te_3$ and $Sb_2Te_3$ have almost three times higher carrier density than $Bi_2Te_3$ and more than seven times higher than $Bi_2Te_2Se$ across all temperatures investigated. This plot confirms that Sb-TI samples are significantly more conductive than all other TI samples.

In **Figure 6**, we fitted $T_1$ from a saturation-recovery experiment for the phase separated $Bi_2Te_2S$ at 355 K and room temperature for the central versus shoulder peaks and found significantly different $T_1$ values for each peak. We note that the shoulder peak relaxes 1.8 times faster than the central peak at both temperatures. This is an interesting result because by increasing the temperature, an even partial annealing effect would have been expected to show a gradual change in $T_1$ as opposed to the difference observed between the shoulder and central peaks. The PXRD patterns (**Figure 1**) do not provide evidence of any damage to the bulk, suggesting that the observed positive frequency shift peak is likely primarily a phase separation effect. $^{125}$Te NMR line shape measurements in Bi$_2$Te$_2$S at 355K, as shown in **Figure 6**a, provide experimental evidence about the existence of two distinct regions (phase separation, see also **Figure 1**c). Furthermore, the NMR shoulder peak characteristics (frequency shift and linewidth) evolve with temperature in a way similar to "pure" Bi$_2$Te$_3$, as shown in **Figure 3**, which further supports the view that this material consists of two coexisting phases.

## 4. Conclusions

A variable temperature $^{125}$Te NMR study has been performed on Bi$_2$Te$_3$, Sb$_2$Te$_3$, Bi$_{0.5}$Sb$_{1.5}$Te$_3$, Bi$_2$Te$_2$Se and Bi$_2$Te$_2$S, revealing fundamental differences between these five chalcogenides which originate from the presence of different defect content in the tetradymite matrix. Asymmetric static lineshapes due to a distribution of Knight shifts in various domains of the sample, arising from the presence of antisite defects as the substitutional component (Se, S or Sb in Bi$_2$Te$_3$ matrix) concentration increases. Relaxation studies show that the binary TIs exhibit different characteristics compared to the ternary sample suggesting a range of charge carrier densities due to an extensive distribution of defects. A lower thermal activation energy and a lower Korringa product accompanied by a stronger negative frequency shift suggest a more conductive bulk profile for Bi$_{0.5}$Sb$_{1.5}$Te$_3$ than the two parent binary TIs. On the other hand, Bi$_2$Te$_2$S and the ordered Bi$_2$Te$_2$Se maintain the lower values of $1/(T_1 \cdot T^{0.5}) \propto N$ featuring a more insulating behavior across the entire temperature range. This result is consistent with recent magneto-transport studies[1-6] of the bulk state resistivity; this is consistent with PXRD and NMR results in **Table 1**. We also provide an experimental evidence about the existence of two distinct regions in the Bi$_2$Te$_2$S composition. Due to the high sensitivity of the NMR technique to sample defects, homogeneity and carrier concentration, this study further supports previous reports asserting that the ordered ternary compound Bi$_2$Te$_2$Se is the best TI material among those investigated herein. This is because the bulk state remains insulating across the entire temperature range, allowing the surface states to dominate the material's conductivity.

## 5. Experimental Section

*Materials*: $Sb_2Te_3$ and $Bi_2Te_3$ ingots were purchased from Alfa Aesar, $Bi_{0.5}Sb_{1.5}Te_3$ from Sigma-Aldrich and used without further purification or recrystallization. According to the manufacturers, samples are grown in solid state reactions by reacting elemental metals in a high temperature furnace to form ingots. $Bi_2Te_2Se$ and $Bi_2Te_2S$ ingots were synthesized by mixing the appropriate ratios of elemental Bi, Te, and Se/S. Tubes were sealed under high vacuum (~$10^{-4}$ Torr), heated to 800 °C over 15 hours, kept at 800 °C for 10 hours, and cooled to room temperature over 15 hours.

*Structural, electrical and Nuclear Magnetic Resonance characterization*: Powder *X-ray* diffraction (PXRD) analysis was carried out on a Panalytical X'Pert Pro *X-ray* Powder Diffractometer with *Cu Kα* radiation ($\lambda$ = 1.54050 Å). Particle sizes reported for the samples are average particle sizes determined from PXRD data. For the commercial compositions, $Bi_2Te_3$, $Sb_2Te_3$ and $Bi_{0.5}Sb_{1.5}Te_3$, a range of carrier concentrations is given based on literature data. For the $Bi_2Te_2Se$ and $Bi_2Te_2S$ compositions, carrier concentrations were determined using measurements of Hall coefficients at room temperature with a home-built system in an applied magnetic field. For NMR studies, the ingots were ground by mortar and pestle. PXRD results as shown in Table 1 (lattice parameters) and in Figure 1b (indexing powder patterns) indicate that the samples remain crystalline after the mortar and pestle process. The PXRD studies indicate that the bulk is damage free. Damage to the edges probably exists, but is of no consequence here because we are probing the bulk of these materials (our powders are micrometer sized, so the surface-to-volume effects are negligible). $^{125}$Te NMR data were acquired on static powder samples with a Bruker DSX-300 spectrometer operating at 94.79

MHz using a standard Bruker X-nucleus wideline probe with a 5-mm solenoid coil. The $^{125}$Te $\pi/2$ pulse width in the wideline probe was 4 $\mu$s. Spectral data were acquired using a spin-echo sequence [$\pi/2$) x – $\tau$ – $\pi$) y - acquire]. The echo delay, $\tau$, was set to 20 μs. The spin-echo sequence is useful in minimizing pulse ringdown effects. In order to acquire the full $^{125}$Te NMR spectrum of the samples, we used the variable offset cumulative spectra technique.[66] Data for determining the spin-lattice relaxation times ($T_1$) were acquired with the saturation-recovery technique. The $^{125}$Te chemical shift scale was calibrated using the unified $\Xi$ scale, relating the nuclear shift to the $^1$H resonance of dilute tetramethylsilane in CDCl$_3$ at a frequency of 300.13 MHz.[67]

**Acknowledgments**

This research was supported by the Defense Advanced Research Project Agency (DARPA), Award No. N66001-12-1-4034. We acknowledge the use of instruments at the Molecular Instrumentation Center (MIC) facility at UCLA.

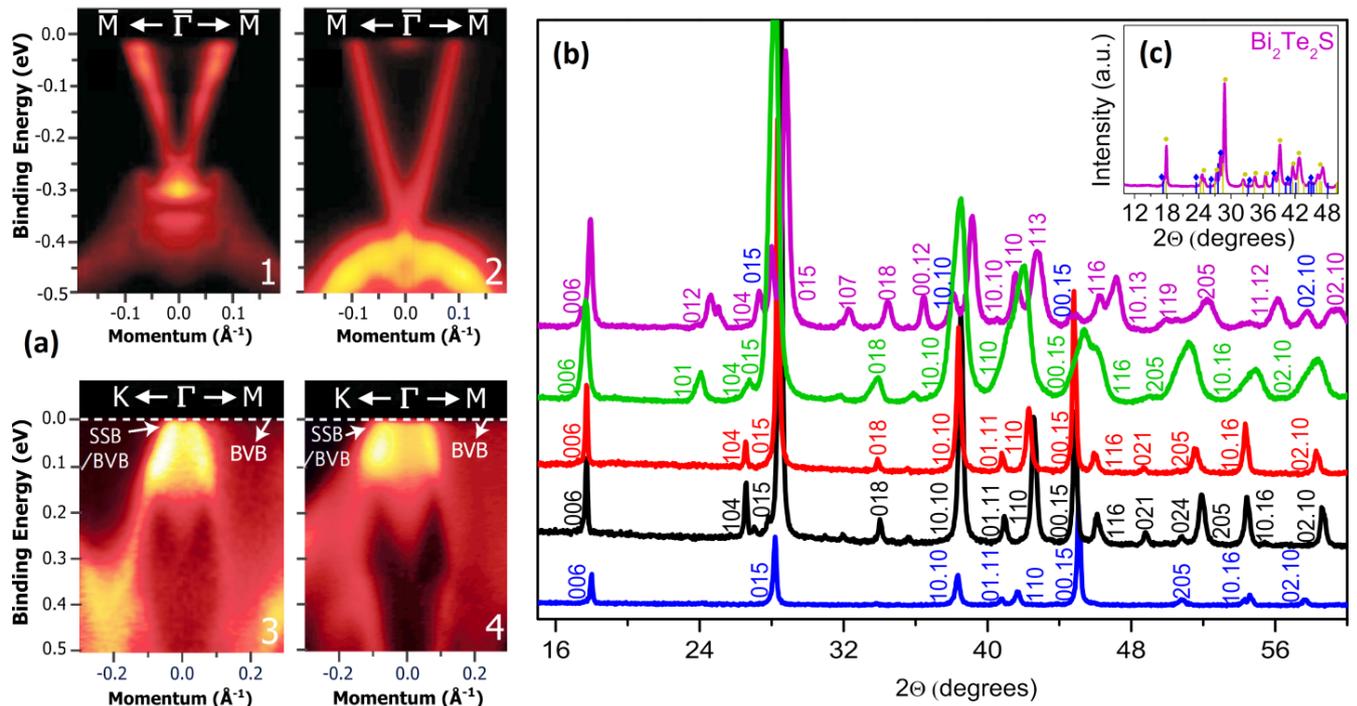

**Figure 1.** Comparison of ARPES results (a) of the band structures of (1) $Bi_2Te_3$ and (2) $Bi_2Te_2Se$, (3) $Bi_{0.5}Sb_{1.5}Te_3$, (4) $Sb_2Te_3$, along the K-Γ-M and M-Γ-M symmetry momentum directions (a). PXRD patterns from mortar and pestle powders $Bi_2Te_3$ (blue), $Bi_{0.5}Sb_{1.5}Te_3$ (red), $Sb_2Te_3$ (black), $Bi_2Te_2S$ (purple) and $Bi_2Te_2S$ (green) at room temperature (b). PXRD patterns of all the samples confirm their crystallinity. $Bi_2Te_2S$ is a macroscopically phase separated TI with one of the phases being S-poor $Bi_2Te_3$ and the second one, $Bi_2Te_{2+\delta}S_{1-\delta}$ (c). [ARPES images 1 and 2 (top row) adapted from Ref. 5; 3 and 4 (bottom row) adapted from Ref. 40.]

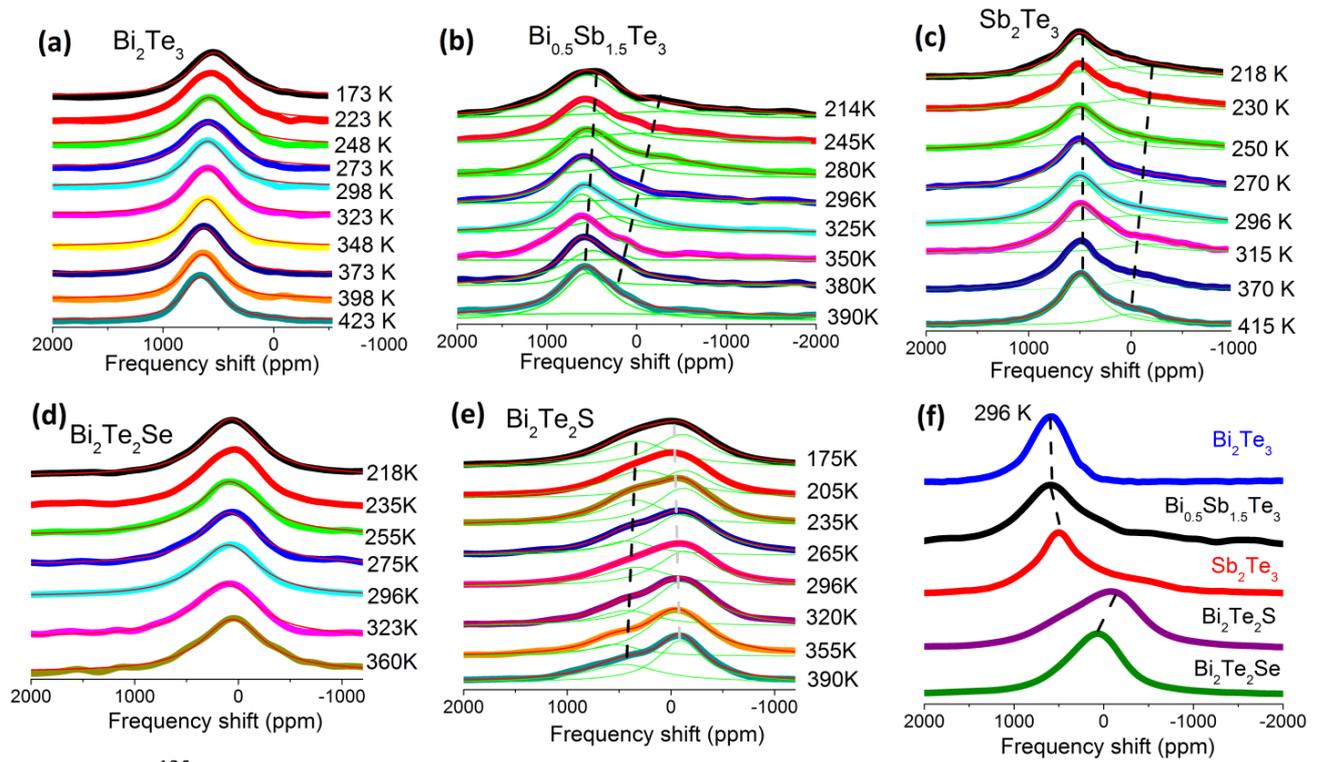

**Figure 2.** $^{125}$Te NMR powder spectra for (a) $Bi_2Te_3$, (b) $Bi_{0.5}Sb_{1.5}Te_3$, (c) $Sb_2Te_3$, (d) $Bi_2Te_2Se$ and (e) $Bi_2Te_2S$ as a function of temperature. The red thin lines are multi-component best fits, and the green thin lines are the individual components. A comparative plot (d) of NMR powder spectra at 296 K indicates a negative frequency shift as the Sb concentration increases in the case of the *p*-type materials and a positive frequency shift for the *n*-type TIs. The $Bi_2Te_2S$ as a phase separated ($Bi_2Te_3$-$Bi_2Te_{1.6}S_{1.4}$) material reveals a left shoulder peak (e). The left shoulder peak is attributed to the $Bi_2Te_3$-rich phase since it has equal frequency shift with $Bi_2Te_3$ (f).

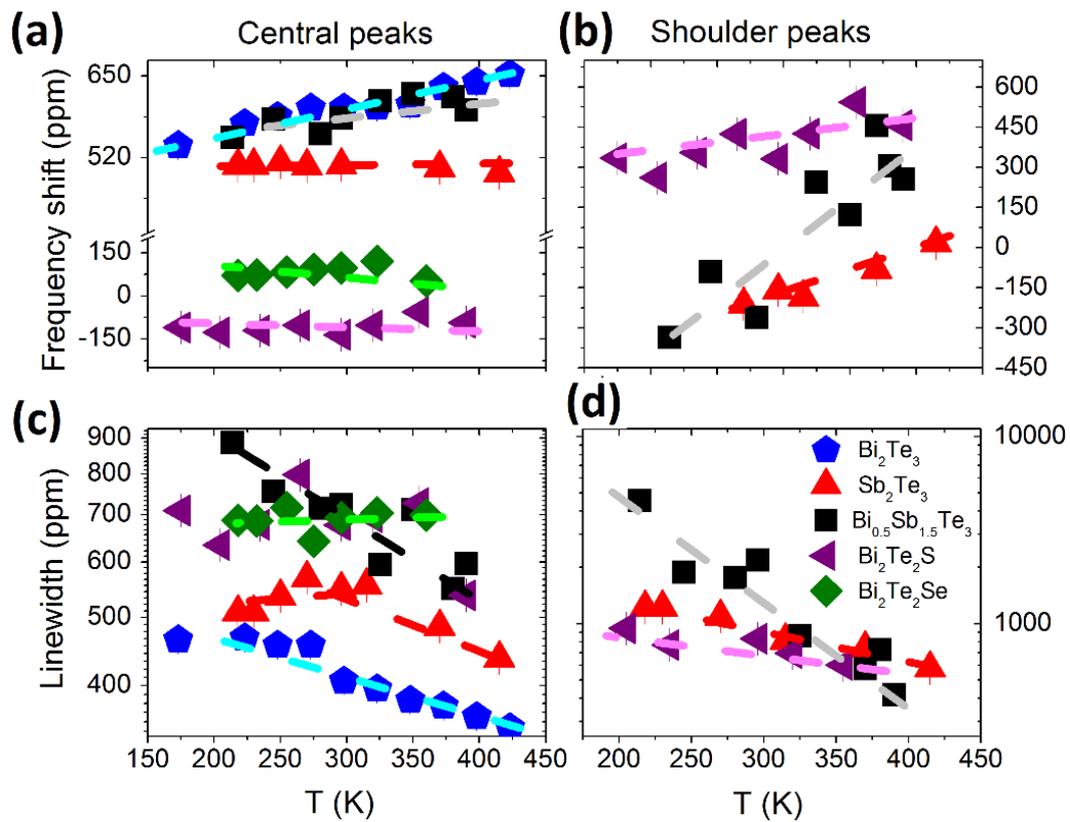

**Figure 3.** Temperature dependence of the frequency shift (a,b) and linewidth (c,d) for Bi$_2$Te$_3$ (●), Sb$_2$Te$_3$ (▲, central and shoulder peak), Bi$_{0.5}$Sb$_{1.5}$Te$_3$ (■, central and shoulder peak), Bi$_2$Te$_2$Se (♦) and Bi$_2$Te$_2$S (◄, central and shoulder peak).

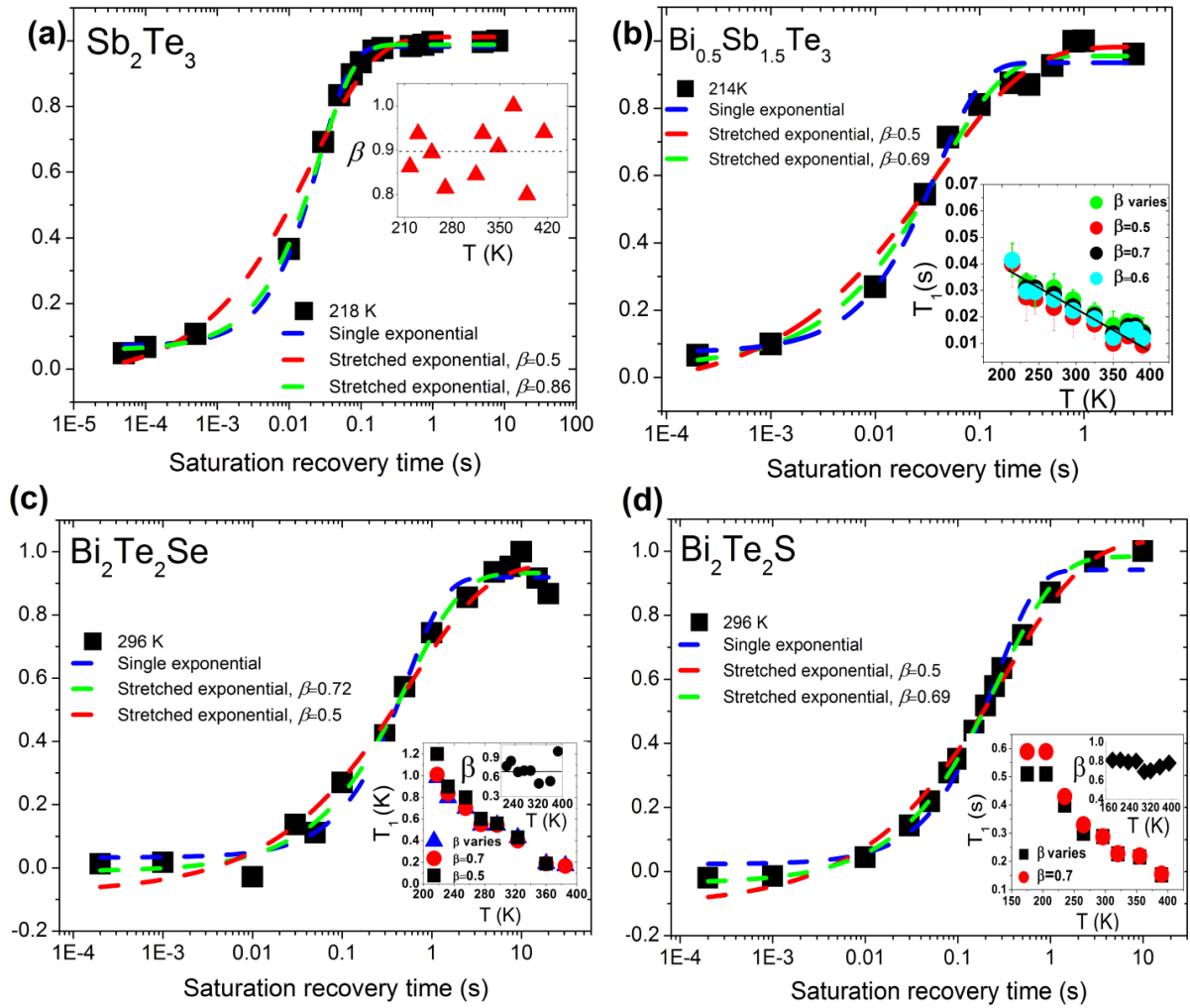

**Figure 4.** Spin-lattice saturation recovery ($T_1$) relaxation data of the entire resonance for (a) $Sb_2Te_3$ whereas the inset presents the temperature dependence of $\beta$. (b) $Bi_{0.5}Sb_{1.5}Te_3$ and the inset reveals that the $\beta$ does not affect the overall trend of $T_1$. $Bi_2Te_2Se$ at 296K and the $T_1$ $\beta$-independence across the temperature range (c) and (d) $Bi_2Te_2S$ (296 K). The blue dashed line shows a fit of a single exponential ($\beta$=1), the red ($\beta$=0.5) and green (by allowing the $\beta$ parameter to vary) dashed lines the fits with stretched exponential.

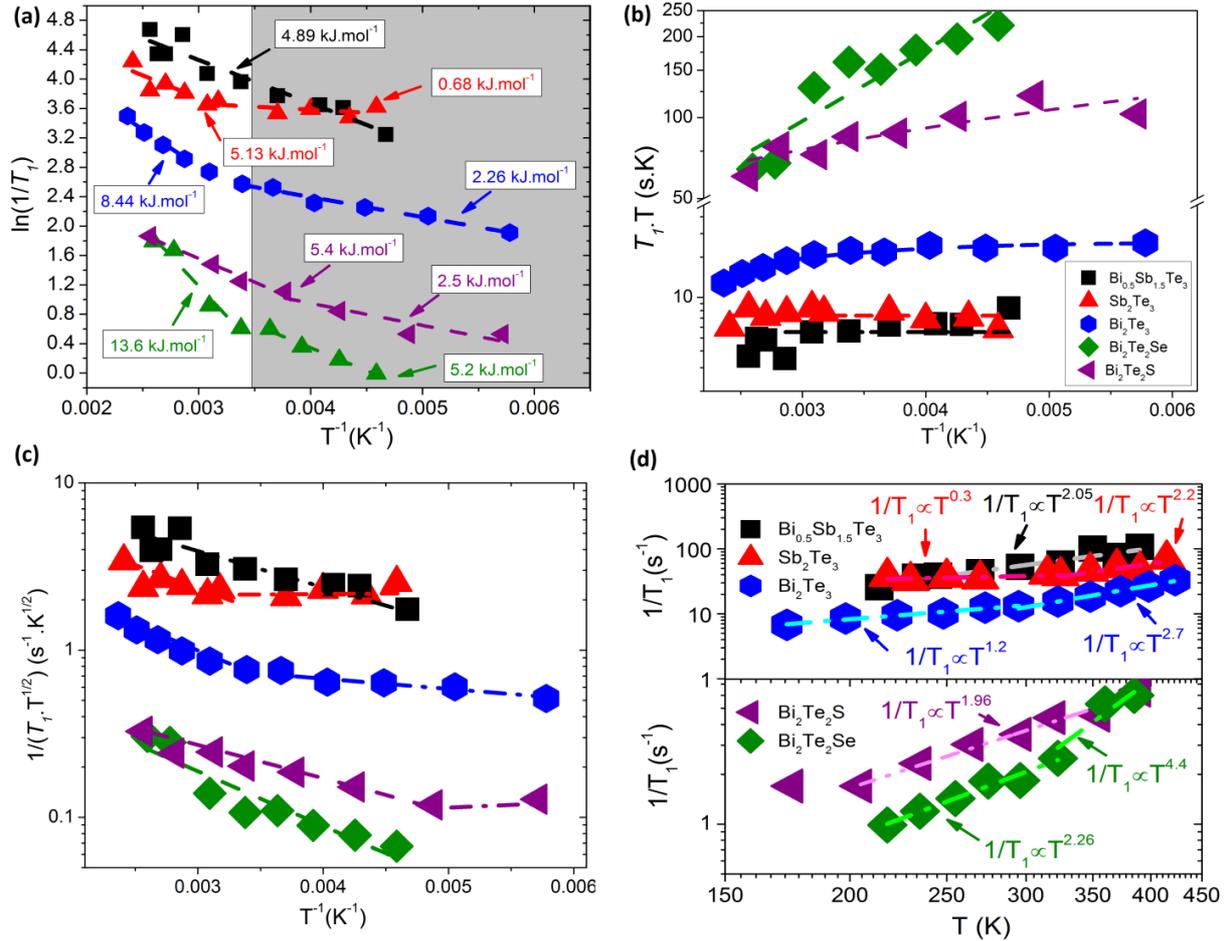

**Figure 5.** The natural logarithm of the $^{125}$Te spin-lattice relaxation rate in the case of $Bi_{0.5}Sb_{1.5}Te_3$ (■), $Sb_2Te_3$ (▲), $Bi_2Te_3$ (●), $Bi_2Te_2S$ (◀) and $Bi_2Te_2S$ (◆) as a function of the inverse temperature (a). Below 270 K, a second mechanism dominates the spin-lattice relaxation. Comparison of the Korringa product (b) of the $^{125}$Te spin-lattice relaxation time with the temperature as a function of the inverse temperature for $Bi_{0.5}Sb_{1.5}Te_3$ (■), $Sb_2Te_3$ (▲), $Bi_2Te_3$ (●), $Bi_2Te_2S$ (◀) and $Bi_2Te_2Se$ (◆) m&p samples. A semi-logarithmic plot for $1/(T_1 \cdot T^{0.5})$ vs inverse temperature of the three TIs. The quantity $1/(T_1 \cdot T^{0.5})$ is proportional to charge carrier concentration (see text and Table I). The temperature dependence of the spin-lattice relaxation rate for all the samples (d). The dashed lines show power-law fit $(1/T_1 \propto T^n)$ of relaxation rates. The power law fit gives a slow temperature dependence of the form $1/T_1 \propto T^{0.3}$ only for the $Sb_2Te_3$ and a $1/T_1 \propto T^{1.2}$ for the $Bi_2Te_3$ below 270 K.

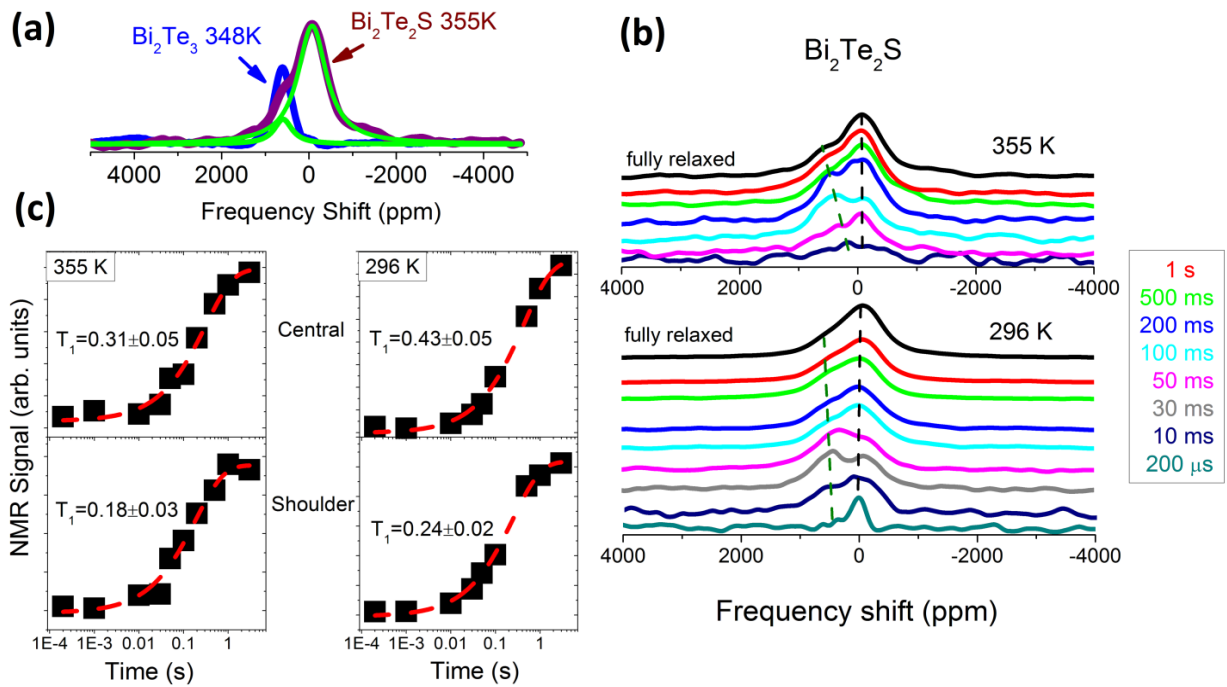

**Figure 6.** The $^{125}$Te NMR spectra for the phase separated Bi$_2$Te$_2$S (purple line) at 355 K and Bi$_2$Te$_3$ (blue line) at 348 K. The green thin lines are the individual components of a two-Lorentzian fitting (a). The shoulder peak at more positive shifts of Bi$_2$Te$_2$S corresponds to the S-poor Bi$_2$Te$_3$ phase which is in agreement with the PXRD pattern analysis as shown in Figure 1c. Comparison of the $^{125}$Te spectrum for the phase-separated Bi$_2$Te$_2$S at 355 K and 296 K obtained in the saturation recovery experiment after saturation of 200 μs to 3 s (b). Relaxation recovery is not uniform across the resonance. The regions at more positive shifts (shoulder peak) relax more quickly (c) than those at more negative shifts (central peak). The red line are fitted to a stretched exponential saturation-recovery model with the $\beta$=0.7.

**Table 1.** Structural and transport parameters obtained from PXRD and NMR analysis. For the $Bi_2Te_2S$ the lattice parameters are referred to the S-poor and S-rich phases. The carrier concentration ($N$) for $Bi_2Te_3$, $Bi_{0.5}Sb_{1.5}Te_3$ and $Sb_2Te_3$ are based on literature data. The $N$ values for the $Bi_2Te_2Se$ and $Bi_2Te_2S$ were determined from Hall measurements of the present study.

| Composition | Lattice parameters [Å] | | | | Volume [Å$^3$] | | Activation energy [kJ/mol] | | Carrier concentration [cm$^{-3}$] | Ref. | $1/(T_1 \cdot T^{0.5}) \propto N$ [s$^{-1}$·K$^{1/2}$] |
|---|---|---|---|---|---|---|---|---|---|---|---|
| | $a$ | | $c$ | | $V$ | | Low T[a] | High T[b] | $N$ | | |
| $Bi_2Te_3$ | 4.322 | | 29.535 | | 477.888 | | 2.26 | 8.44 | $(1-5) \cdot 10^{19}$ | c | 0.76 |
| $Bi_{0.5}Sb_{1.5}Te_3$ | 4.268 | | 30.134 | | 473.512 | | 4.89 | | $(4-7) \cdot 10^{19}$ | d | 3.05 |
| $Sb_2Te_3$ | 4.239 | | 30.034 | | 467.542 | | 0.68 | 5.63 | $(3-8) \cdot 10^{19}$ | e | 2.83 |
| $Bi_2Te_2Se$ | 4.292 | | 30.111 | | 480.580 | | 5.20 | 13.60 | $8.55 \cdot 10^{18}$ | f | 0.10 |
| $Bi_2Te_2S$ | 4.349 | 4.226 | 30.487 | 29.500 | 499.300 | 456.210 | 2.50 | 5.40 | $1.82 \cdot 10^{19}$ | f | 0.20 |

[a] Temperature range below 270 K; [b] Temperature range above 270 K; [c] Reference [10,11, 34, 35, 43]; [d] Reference [10]; [e] Reference [10, 21, 23, 25, 35]; [f] Current study.